\documentclass[aps,prl,showpacs,preprint,floatfix]{revtex4}\usepackage{latexsym}
\usepackage{latexsym}
\usepackage{amsfonts}
\usepackage{amssymb}
\usepackage{amsmath}
\usepackage{graphicx}

\begin{document}

\title{Frequency converter based on nanoscale MgO magnetic tunnel junctions}

\author{B. Georges, J. Grollier, V. Cros, B. Marcilhac, D.-G. Cr\'et\'e, J.-C. Mage, A. Fert}
\affiliation{Unit\'e Mixte de Physique CNRS/Thales and Universit\'e Paris Sud 11, 1 Avenue A. Fresnel, 91767 Palaiseau Cedex}
\author{A. Fukushima, H. Kubota, K. Yakushijin, S. Yuasa, K. Ando}
\affiliation{National Institute of Advanced Industrial Science and Technology (AIST) 1-1-1 Umezono, Tsukuba, Ibaraki 305-8568, Japan}

\date{\today}

\begin{abstract}
We observe both dc voltage rectification and frequency conversion that occur when a reference microwave current is injected to a MgO based magnetic tunnel junction (MTJ). The rectification that is spin-transfer torque dependent is observed when the frequency of the input microwave current coincides with the resonance frequency of the magnetization of the active layer. In addition, we demonstrate that frequency conversion is the result of amplitude modulation between the reference signal and the resistance of the MTJ that is fluctuating at the resonance frequency of the magnetization of the active layer. 
\end{abstract}

\maketitle

MgO-based magnetic tunnel junctions (MTJs) are extensively studied because of their implementation in hard disk reading heads or as bit cells in magnetic memories MRAM. Recent improvements in their fabrication process resulting in low resistive insulating barrier has made possible the observation of spin-transfer torque phenomena \cite{Yuasa, Parkin}. The transfer of spin angular momentum from a spin-polarized current to a ferromagnet enables to manipulate magnetization without magnetic field \cite{SloncBerger}. This effect can be used for example to induce steady magnetization precessions by injection of a dc current. These magnetic dynamical regimes are converted into a microwave electrical signal through the magneto-resistive effect \cite{Tsoi,Kiselev2003,Rippard2004, Nazarov, Deac}. Spin transfer nano-oscillators (STNOs) offer high potentialities for a new generation of nanoscale high frequency synthetizer for telecommunication applications. Standard microwave features such as the frequency modulation of  the STNO output signal has also been evidenced when a low frequency ac current is added to the dc current \cite{PufallModulation}. Moreover, for an excitation frequency much closer to the natural STNO one, frequency locking occurs \cite{RippardPRLinjection, GeorgesPRL}. Analytical calculation of such phenomena has been extended to the description of the phase locking in arrays of coupled STNOs \cite{GeorgesAPL}. In addition, the injection of a pure high frequency signal at the resonance frequency of the active magnetic layer is known to generate a dc rectified voltage across the device 
 
In this Letter, we demonstrate the frequency conversion of an input microwave signal obtained by the mixing with the ferromagnetic resonance frequency $f_{FMR}$ of the free magnetic layer in a MgO-based MTJ, not biased with dc current. When the input signal frequency $f_{rf}$ coincides with $f_{FMR}$, a dc rectified voltage is measured across the device. For $f_{rf} \neq f_{FMR}$, microwave spectra are characterized by two modulation peaks at respectively $\left|f_{rf} - f_{FMR}\right|$ and $f_{rf} + f_{FMR}$. This proves that, potentially, magneto-resistive devices could be used as a powerless nano-scale microwave frequency mixer.

The samples used are magnetic tunnel junctions composed of PtMn (15)/ CoFe (2.5) / Ru (0.85) / CoFeB (3) / MgO (1.075) / CoFeB (2) (nm) patterned into an elliptical shape of dimension 170x70 $nm^{2}$ \cite{Nagamine}. The tunnel magneto-resistance ratio (TMR) is 100$\%$ and the RA product is 0.85 $\Omega$.$\mu$$m^{2}$ for the parallel (P) magnetization configuration at room temperature. An in-plane magnetic field $H$ (between 150 to 500 Oe) is applied along the hard axis of the ellipse in order to increase the amplitude of the measured microwave signals.

\begin{figure}[h]
   \centering
    \includegraphics[width=8.5 cm]{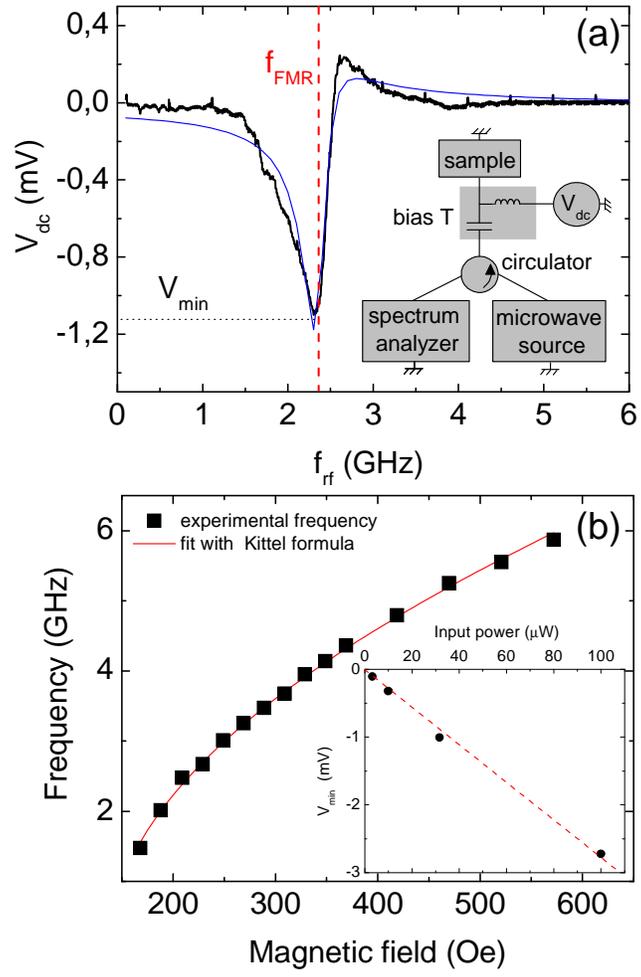}
     \caption{Georges \textit{et al.}}
\label{fig1}
\end{figure}

A microwave current delivered by an external source is injected into the sample through the capacitive branch of a bias tee via a 2-4 GHz circulator  as shown in the inset of Fig.\ref{fig1} (a). The dc voltage across the MTJ is measured at the inductive output of the bias tee. The microwave part of the output signal is recorded onto a spectrum analyzer connected to the third circulator port. The frequency of the microwave input signal $f_{rf}$ is swept from 100 MHz up to 6 GHz. The rectified dc and microwave output voltages have been recorded for four values of the input power: $P$ = -25, -20, -15 and -10 dBm. The actual amplitude of the microwave current passing through the sample depends on the frequency because of the mismatch impedance and the use of the circulator. We estimate that the maximum injected current obtained for $P$ = -10 dBm is about 0.4 mA at $f_{rf}$ = 3 GHz.

In Fig.\ref{fig1} (a), we show a typical rectified dc voltage measured as function of $f_{rf}$ for $H$ = 210 Oe and $P$ = -15 dBm. An asymmetric negative peak with a minimum of rectification $V_{min}$ = -1.1 mV is observed at 2.32 GHz. As already described in \cite{Sankey, Kubota}, this peak corresponds to the electrical response of the spin-transfer induced resonant excitation of the magnetization of the free layer. The calculation of the rectified dc voltage as function of $f_{rf}$ can be obtained from the Landau-Lifshitz-Gilbert including the spin transfer torques. It corresponds to the dc component of $I_{rf}cos(2\pi f_{hf}t) \Delta R(I_{rf}, f_{rf})$, where $\Delta R(I_{rf}, f_{rf})$ is the frequency dependent resistance variation when the magnetic moment of the free layer is harmonically excited at its resonance frequency by a microwave current. As $\Delta R(I_{rf}, f_{rf})$ is proportional to $I_{rf}$, it comes out that the rectified voltage goes as $I_{rf}^{2}$. We confirm this behavior by measuring a linear variation of the measured minimum rectified voltage $V_{min}$ as a function of $P$ (see inset of fig.\ref{fig1}(b)). Moreover, the lineshape of $V_{dc}$($f_{rf}$) is composed of symmetric and antisymmetric lorentzian attributed respectively to the Slonczewski-torque and the field-like torque (see plain line in Fig.\ref{fig1}). By fitting the experimental lineshape of $V_{dc}$($f_{rf}$) at different magnetic fields, we extract the dependence of the resonance frequency $f_{FMR}$ with $H$, that can be fitted using the Kittel formula (see Fig.\ref{fig1}(b)). For such geometry, the resonance frequency is given by $f_{FMR} = \frac{\gamma}{2\pi} \sqrt{\left( H - H_{an} \right) \left(H+4\pi M_s\right)}$, where $\gamma$ = 28$.10^{-4}$ GHz/Oe is the gyromagnetic ratio, $H_{an}$ is the uniaxial anisotropy field and $4\pi M_s$ is the effective magnetization. The best fitting parameters are $H_{an}$ = 137 Oe and $4\pi M_s$ = 9900 Oe. Then, from the spin-diode measurements, we have access to the resonance frequency of the active layer at any field.

\begin{figure}[h]
   \centering
    \includegraphics[width=8.5 cm]{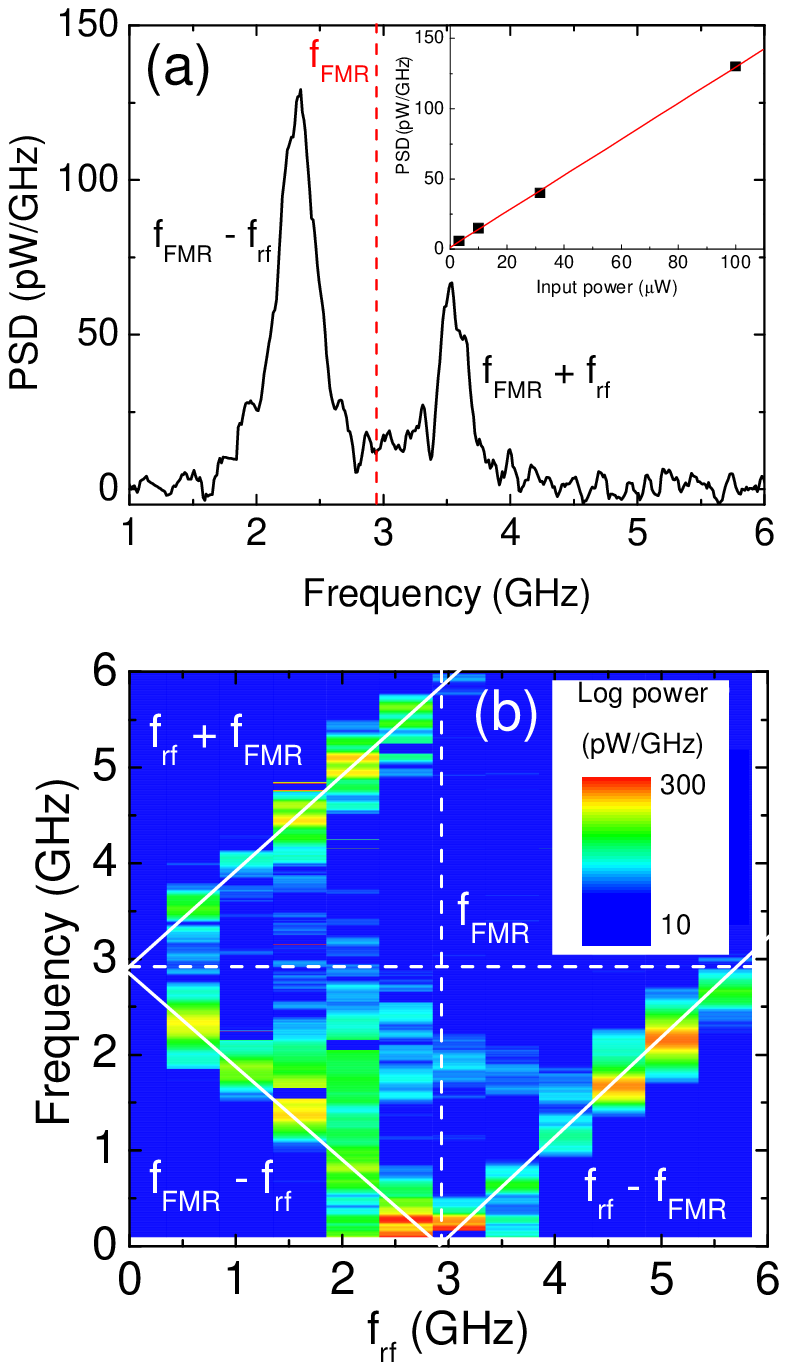}
     \caption{Georges \textit{et al.}}
\label{fig2}
\end{figure}

We now turn to the description of the microwave output signals. In Fig.\ref{fig2}(a), we show the power spectra obtained at $H$ = 250 Oe with $f_{rf}$ =  0.6 GHz and P = -10 dBm (no dc current is applied). Two peaks at 2.34 and 3.54 GHz are detected at frequencies corresponding respectively to $f_{FMR} - f_{rf}$ and $f_{rf} + f_{FMR}$ (the position of the resonance frequency $f_{FMR}$ = 2.94 GHz is indicated by the vertical dashed line). In Fig.\ref{fig2}(b), we present the evolution of the peak positions and amplitudes (in color scale) with $f_{rf}$. As $f_{rf}$ increases up to $f_{FMR}$, the two peaks follows their respective branches $f_{FMR} - f_{rf}$ and $f_{FMR} + f_{rf}$. Above $f_{FMR}$, the lowest frequency branch follows $f_{rf} - f_{FMR}$. These peaks are characteristic of the modulation between two signals with different frequencies. 

In our experiment, no dc current is applied and the resonance frequency $f_{FMR}$ is weakly dependent on the rf current over the range $\pm$ {0.4} mA, thus frequency modulation is negligible in this case. Therefore the observed modulation phenomena is associated to amplitude modulation (AM). Indeed, if we assume that the magnetization is fluctuating at its resonance frequency, the MTJ resistance has an oscillating component at $f_{FMR}$: $R(t, T) = R_0 + \Delta R(T) cos(2\pi f_{FMR}t)$, where $R_0$ is the nominal resistance. The amplitude of the resistance variation $\Delta R(T)$, related to the amplitude of the magnetization fluctuations, is only determined by the temperature $T$. When injecting a microwave signal at the frequency $f_{rf}$, the resulting output voltage is:
\begin{eqnarray} 
V_{output} =  R(t, T) I_{rf}cos(2\pi f_{rf}t) \nonumber \\
V_{output} =  R_0 I_{rf}cos(2\pi f_{rf}t) \nonumber \\
+ \frac{\Delta R(T) I_{rf}}{2} \{cos\left[2\pi \left(f_{FMR}-f_{rf}\right)t)\right] + cos\left[2\pi \left(f_{FMR}+f_{rf}\right)t)\right]\},
\label{voltage}
\end{eqnarray} 

that corresponds to an equation of amplitude modulation. In case of non-linear effects that would induce a dependence of $f_{FMR}$ with the applied current, amplitude modulation but also frequency modulation will occur \cite{PufallModulation}.

The last two terms of Eq.\ref{voltage} indicates that the output voltage has a component at $\left|f_{FMR}-f_{rf}\right|$ and another one at $f_{FMR}+ f_{rf}$. It is worth emphasizing that their amplitudes are proportional to $I_{hf}$ and not to $I_{rf}^{2}$ like in the resonant case, i.e. $f_{FMR} = f_{rf}$. This difference is due to the fact that the resistance variation associated to the magnetization fluctuations should not depend on $I_{rf}$. In the inset of Fig.\ref{fig2}(a), we display the evolution of the $f_{FMR}-f_{hf}$ - modulated peak PSD ($\propto V_{output}^{2}$) with the source power ($\propto I_{rf}^2$), obtained for $H$ = 250 Oe, and $f_{rf}$ = 0.6 GHz. We observe a linear response of the output voltage with the microwave current, confirming that, in this regime, the amplitude of the magnetization motion is independent on $I_{hf}$. 

To summarize, we have measured simultaneously the dc and microwave output voltages across a MgO-based MTJ while injecting only a microwave current. For an injection frequency equal to the resonance frequency of the magnetization of the active layer, a rectified dc voltage occurs through the TMR effect at the MTJ. For this resonant excitation case, the amplitude of the magnetization motion, then the resistance variation,  is proportional to the microwave current amplitude, leading to a rectified voltage proportional to $I_{rf}^2$. For any other frequency, the microwave output spectra show modulated peaks between the resonance frequency and the input frequency. As, in this case, the magnetization is at the thermal equilibrium, its motion amplitude, and then the resistance variation, only depends on the temperature, and not on the microwave current, leading to an microwave output voltage proportional only to $I_{rf}$. These results demonstrate that magneto-resistive devices might be used as nano-sized microwave mixer operating at high frequencies.

B. G. is supported by a PhD grant from the DGA. We acknowledge Canon-Anelva Corp. for the fabrication of MTJ films. This work is partially supported by the CNRS and ANR agency (NANOMASER PNANO-06-067-04).

\newpage

\textbf{Figure captions}\\

Figure 1.(a) Power spectral density (PSD) showing the modulation phenomena measured for $f_{rf}$ = 0.6 GHz, $P$ = -10 dBm and $H$ = 250 Oe. The position of the resonance frequency $f_{FMR}$ is shown by the vertical dashed line. \textit{Inset:} Dependence of the $f_{FMR} - f_{rf}$ peak PSD as function of the input power $P$ for similar experimental conditions. (b) Color map showing the evolution of the modulated peaks as function of the input frequency $f_{rf}$, for $P$ = -10 dBm and $H$ = 250 Oe. The color scale is the measured PSD.\\

Figure 2.(a) Power spectral density (PSD) showing the modulation phenomena measured for $f_{rf}$ = 0.6 GHz, $P$ = -10 dBm and $H$ = 250 Oe. The position of the resonance frequency $f_{FMR}$ is shown by the vertical dashed line. \textit{Inset:} Dependence of the $f_{FMR} - f_{rf}$ peak PSD as function of the input power $P$ for similar experimental conditions. (b) Color map showing the evolution of the modulated peaks as function of the input frequency $f_{rf}$, for $P$ = -10 dBm and $H$ = 250 Oe. The color scale is the measured PSD.

\newpage

\newpage

\end{document}